\documentclass[11pt,a4paper,twocolumn]{book}
\usepackage{jnpcs}
\usepackage{graphics}
\usepackage{epsfig}
\usepackage{epstopdf}

\ntitle{A model solution of the generalized Langevin equation: \\ Emergence and Breaking of Time-Scale Invariance\\ in Single-Particle Dynamics of Liquids}

\nauthor{Anatolii~V.~Mokshin$^{1,2}$ and Bulat~N.~Galimzyanov$^{1,2}$}

\naddress{$^1$Institute of Physics, Kazan Federal University,
      \it 420008 Kazan, Russia \\
   $^2$L.D. Landau Institute for Theoretical Physics, Russian Academy of Sciences,
   \it 117940 Moscow, Russia \\
   {\small \rm E-mail:  anatolii.mokshin@mail.ru}}

\ndata{\today}

\nabstract{It is shown  that the solution of generalized Langevin equation
can be obtained on the basis of simple comparison of the time scale for the
velocity autocorrelation function of a particle (atom, molecule) and of the time scale for the corresponding memory
function. The result expression for the velocity autocorrelation function contains dependence on
the non-Markovity parameter, which allows one to take
into account memory effects of the investigated phenomena. It is demonstrated for the cases of liquid tin and liquid lithium that the obtained expression for the velocity autocorrelation function is in a good agreement with the molecular dynamics simulation results.
\begin{flushright}
{\it Dedicated to the memory of Prof.~Renat M. Yulmetyev}
\end{flushright}
}

\nkeywords{memory effects, stochastic processes, non-linear
dynamics, non-Markovian processes, velocity autocorrelation
function, molecular dynamics}

\nPACSnumbers{66.10.Cb,61.20.-p,05.40.-a}

\begin{document}

\DeclareGraphicsExtensions{.jpg,.pdf,.mps,.png} \firstpage{1}
\nlpage{2} \nvolume{4} \nnumber{1} \nyear{2001}
\def\nfpage{\thepage}
\thispagestyle{myheadings} \npcstitle

Relaxation processes in a complex system can be characterized by the pronounced memory effects, which are manifested in the non-exponential decay or oscillatory behavior of the time correlation functions (TCF's) for the corresponding dynamical variables~\cite{Mokshin/Yulmetyev/Hanggi_PRL_2005,Sibani_Complex_systems}. Hence, one can reasonably assume that the direct accounting for the memory effects could simply the theoretical description of the system behavior. The convenient way to examine this is to consider a system, in which the origin of the memory effects is well studied. As an example of such the physical system one can take a high-density (viscous) liquid, a supercooled liquid or/and a glass~\cite{Stanley_1971}, where the memory effects appear in single-particle dynamics as well as in collective particle dynamics~\cite{Boon,Hansen,Hanggi}.

From theoretical point of view, the convenient way to take these
effects into account most adequately is to use in the description
the so-called memory function formalism~\cite{Zwanzig_book}, which
is associated with the projection operators technique of Zwanzing
and Mori~\cite{Zwanzig_1961,Mori_1965} as well as with the recurrent
relations method suggested by Lee~\cite{Lee}. Remarkably, the memory
function formalism allows one to represent the equation of the
motion for a variable (originally, for the velocity of a particle in
liquid) in the form of a non-Markovian integro-differential
equation, which contains a characteristic component -- a memory
function.  For the case when the velocity of a particle represents
such the variable, the integro-differential equation is known as the
generalized Langevin equation (GLE). Herein, if time behavior of the
memory function is defined, then the solution of the GLE will
determine the evolution of the variable (the velocity) and the
corresponding TCF -- the velocity autocorrelation function (VACF) --
can be computed~\cite{Boon,Hansen}. Nevertheless, although the
technique of projection operators gives a prescription to calculate
the memory function, the direct computations are very difficult to
be realized for real physical
systems~\cite{Lee,Grigolini,Stanislavskii}. In this work, we shall
demonstrate that a solution of GLE can be derived by simple
interpolation of its solutions for the memory-free case, the
strong-memory case and the case with a moderate memory. Further, the
resulted solution will contain the parameter which represents a
quantitative measure of the memory effects.

Let us take the velocity $v_{\alpha}$ of a $\alpha$th particle in
liquid as a dynamical variable. Then, GLE can be written
as~\cite{Hanggi,Zwanzig_1961,Mori_1965}
\begin{equation}
\frac{dv_{\alpha}(t)}{dt}=-\Omega_{1}^{2} \int_{0}^{t}d\tau
M_{1}(t-\tau)v_{\alpha}(\tau) +f(t), \label{gle}
\end{equation}
where $f(t)$ is the random force per unit mass, $M_{1}(t)$ is the
normalized first order memory function, which is related to the
random force $f(t)$ by the second fluctuation-dissipation theorem
\cite{Zwanzig_1961,Mori_1965}, and $\Omega_{1}^{2}$ is the
first-order frequency parameter arising from normalization of
$M_{1}(t)$. Note, it is assumed that $\langle
v_{\alpha}(0)f(t)\rangle=0$. Multiplying Eq.~(\ref{gle}) by
$v_{\alpha}(0)$, taking an appropriate ensemble average
$\langle\ldots\rangle$ and applying further the projection operators
technique, it is possible to obtain a hierarchical chain of
integro-differential non-Markovian equations in terms of TCF's:
\begin{eqnarray}\label{chain}
\frac{dM_{i-1}(t)}{dt}&+&\Omega_{i}^{2}\int_{0}^{t}d\tau
M_{i}(t-\tau)M_{i-1}(\tau)=0,\\
 i&=&1,\;2,\ldots\ . \nonumber
\end{eqnarray}
If VACF is chosen as an initial TCF of this hierarchy,
then GLE will be the first equation (i.e. $i=1$) of this chain~\footnote{Originally, equation~(\ref{gle}) written for the variable-velocity was called as the GLE. Nevertheless, the related integro-differential equation written for the corresponding time correlation function is also mentioned as the GLE in the modern studies~\cite{Zwanzig_book}.}.
In the case,
\[
M_{0}(t)=\frac{\langle v_{\alpha}(0) v_{\alpha}(t)\rangle}{\langle
v_{\alpha}(0)^{2} \rangle}
\]
is VACF; $M_{i}(t)$ is TCF of the corresponding dynamical variable,
which has meaning of the $i$th-order memory
function~\cite{Mokshin_TMP_2015}, whereas $\Omega_{i}^{2}$ is the
$i$th-order frequency parameter. Note that all TCF's of chain
(\ref{chain}) including VACF are normalized to unity for
convenience, i.e.
\[
\lim_{t \to 0}M_{i-1}(t)=1,
\]
\[
i=1,\;2,\; \ldots.
\]
Moreover, applying the operator of Laplace transformation
$\hat{\mathcal{L}}=\int_{0}^{\infty}dt \; e^{-st}[\ldots]$  to equations of chain
(\ref{chain}), one obtains the infinite fraction~\cite{Zwanzig_book}:
\begin{equation}
\widetilde{M}_{0}(s)=\frac{1}{s+\Omega_{1}^{2}\widetilde{M}_{1}(s)}=
\frac{1}{\displaystyle s+\frac{\Omega_{1}^{2}}{\displaystyle
s+\frac{\Omega_{2}^{2}}{\displaystyle
s+\frac{\Omega_{3}^{2}}{\displaystyle s+\ldots}}}}.
\label{fraction}
\end{equation}

It is necessary to note that the $i$th-order memory function $M_{i}(t)$ corresponds to a concrete relaxation
process, the physical meaning of which can be established directly from consideration of the analytical expression for $M_i(t)$.
On the other hand, the squared characteristic time scale $\tau^2$ of the relaxation process can be defined as~\cite{Mokshin/Yulmetyev/Hanggi_PRL_2005}
\begin{equation}  \label{tau}
\tau_{i-1}^2 = \left |  \int_0^{\infty}\; t M_{i-1}(t) \;dt \right | = \left | \lim_{s \to 0} \left ( - \frac{\partial \widetilde{M}_{i-1}(s)}{\partial s}  \right ) \right | .
\end{equation}
\[
i=1,\;2,\; \ldots.
\]
The time scales $\tau_{i-1}$ corresponding to TCF's of chain (\ref{chain})
form the hierarchy, which has the following peculiarity: the
quantity $\tau_{i}$ defines a memory time scale for TCF
$M_{i-1}(t)$, the relaxation time of which is $\tau_{i-1}$. As a
quantitative measure of memory effects for the $i$th relaxation level
it is convenient to use the dimensionless parameter~\cite{Mokshin/Yulmetyev/Hanggi_PRL_2005}
\begin{equation} \label{memory_measure}
\delta_{i}=\frac{\tau_{i-1}^{2}}{\tau_{i}^2}, \ \ \
i=1,\;2,\;\ldots,
\end{equation}
where
$\tau_{i}^2$ is defined by Eq.~(\ref{tau}). This simple criterion
allows one to determine whether the considered process is
characterized by a strong statistical memory, or it has a memoryless behavior. Namely, one has
\begin{equation} \label{eq: limit}
\begin{array}{rl}
\delta \rightarrow 0 & \quad{\textrm{for a strong memory limit,}}\\
\delta \simeq 1 & \quad{\textrm{for a case of moderate memory,}}\\
\delta \rightarrow \infty & \quad{\textrm{for a memory-free limit.}}\\
\end{array} \label{g_np}
\end{equation}

It is remarkable that for the three cases determined by (\ref{eq: limit}) there are known exact solutions of the GLE written for the VACF~\cite{my,Mokshin/Yulmetyev_NJP_2004}:
\begin{equation} \label{first}
\frac{dM_{0}(t)}{dt}+\Omega_{1}^{2}\int_{0}^{t}d\tau \nonumber
M_{1}(t-\tau)M_{0}(\tau)=0, \nonumber
\end{equation}
where the first frequency parameter of a many-particle system (say, for a liquid), where atoms/moleculs interact through a spherical potential $U(r)$, can be written as~\cite{Mokshin/Yulmeyev,Mokshin/Yulmetyev/Khusnutdinov_JETP_2006,Mokshin/Yulmetyev/Khusnutdinoff_JPCM_2007}
\begin{equation}\label{eq: freq_par}
\Omega_{1}^{2}=\frac{4\pi n}{3m}\int_{0}^{\infty} dr \; g(r) r^3 \left[\frac{d}{dr} \left(\frac{1}{r} \frac{d U(r)}{dr} \right )
+\frac{3}{r^2}\frac{dU(r)}{dr}\right].
\end{equation}
Here $n$ is the numerical density, $m$ is the particle mass, and
$g(r)$ is the pair distribution function;  and Eq.~(\ref{first}) is
the first equation of the chain~(\ref{chain}), i.e. at $i=1$. Let us
consider these three cases in detail.

First, one assumes that Eq.~(\ref{first}) describes  the behavior of
the system without memory, i.e. $\tau_{0}^{2} \gg \tau_{1}^2$. For
the case one has $\delta \to \infty$. Here, the memory function
$M_1(t)$ has to decay extremely fast; and, therefore, it can be
taking in the following form:
\begin{equation}\label{dirac}
  M_{1}(t)=2\tau_{1}\delta(t),
\end{equation}
where $\delta(t)$ is the Dirac Delta-function, $\tau_{1}$ is
the time scale of $M_{1}(t)$. By substituting Eq.~(\ref{dirac})
into Eq.~(\ref{first}) and solving the resulted equation, one obtain the VACF $M_{0}(t)$ with ordinary
exponential dependence:
\begin{equation}\label{expon}
  M_{0}(t)=e^{-\Omega_{1}^{2}\tau_{1}t}.
\end{equation}
As known, such the dependence is correct for the velocity correlation function
of the Brownian particle with the relaxation time $\tau_{0}=(\Omega_{1}^{2}\tau_{1})^{-1}=m/\xi_{\beta}$, $m$ and
$\xi_{\beta}$ are the mass and the friction coefficient, correspondingly. As for the self-diffusion phenomena in a liquid, where
particle $(\alpha)$ moving with the velocity $v_{\alpha}(t)$ is
identical to all others, exponential relaxation of the VACF is rather
strongly idealized model~\cite{Resibua}.

Second, one considers the opposite situation appropriate to the
system, the single-particle dynamics of which is characterized by a
strong memory, i.e. $\tau_{0}^{2} \ll \tau_{1}^{2}$. For the case
one obtains from definition~(\ref{memory_measure})  that  $\delta
\to 0$. The most relevant form of the \textit{non-decaying} memory
function can be taken as
\begin{equation}\label{hevisaid}
  M_{1}(t)=H(t)=\left\{
\begin{array}{rl}
1, &t \geq 0\\
0, &t < 0\\
\end{array},
\right.
\end{equation}
where $H(t)$ is the step Heaviside function.
By substituting Eq.~(\ref{hevisaid}) under the convolution integral of Eq.~(\ref{first}), one obtains
\begin{equation}\label{eq: heav}
  \frac{dM_{0}(t)}{dt}=-\Omega_{1}^{2}\int_{0}^{t}d\tau M_{0}(\tau).
\end{equation}
After solving this equation one finds that
\begin{equation}\label{cos}
  M_{0}(t)=\cos(\Omega_{1}t).
\end{equation}
It should be noted that no the characteristic time-scale $\tau_0$ is
included into solution~(\ref{cos}) as well as that $M_{0}(t)$ does
not satisfy the condition of attenuation of correlation at $t \to
\infty$~\cite{Bogol}. Actually, the system with an {\it ideal
memory} ``remembers'' its initial state and returns periodically to
this state, functionally reproducing it with precision.

These two above considered cases are limiting ones. However, it is
known that single-particle dynamics of real systems is characterized
by a memory, albeit the memory is not ideal. Therefore, for the case
one can write that the parameter $\delta$ takes values from the
range  $0<\delta \ll \infty$. There is also the physically correct
solution of Eq.~(\ref{first}) for this region of $\delta$, and the
solution was firstly obtained by Yulmetyev (see
Ref.~\cite{Yulmetyev76}). Let us consider the case, when the time
scales of the initial TCF and of its memory became comparable, i.e.
\[
\delta \simeq 1.
\]
The case can be realized at the time-scale invariance of the
relaxation processes in many-particle
systems~\cite{Mokshin_TMP_2015}.  In addition, if the
time-dependencies of the VACF and of its memory function are
approximately identical, then one can write
\begin{equation}\label{equal}
  M_{1}(t) \simeq M_{0}(t).
\end{equation}
Taking into account relation~(\ref{equal}) and applying Laplace transformation to Eq.~(\ref{first}), we obtain
ordinary quadratic equation:
\begin{equation}\label{square}
  \Omega_{1}^{2}\widetilde{M}_{0}^{2}(s)+s\widetilde{M}_{0}(s)-1=0.
\end{equation}
By solving the last equation and by applying the operator of the
inverse Laplace transformation $\hat{\mathcal{L}}^{-1}$, we find the
VACF
\begin{equation}\label{bessel}
  M_{0}(t)=\frac{1}{\Omega_{1}t}J_{1}(2\Omega_{1}t),
\end{equation}
where $J_{1}(\ldots)$ is the Bessel function of the first kind.
Then, for the squared characteristic time scales of the VACF and of its memory function one obtains from definition (\ref{memory_measure}) and Eq.~(\ref{square}) that
\begin{equation} \label{time}
  \tau_{0}^2=\tau_{1}^2=\frac{1}{2\Omega_{1}^{2}}.
\end{equation}
Thus, the
quantity proportional to the inverse frequency parameter
determines both the squared time scales. Solution (\ref{bessel}) describes the damped
oscillated behavior of the function $M_{0}(t)$. It is worth nothing that the TCF's scenario
is observed frequently in such the physical systems as electron
gas models, linear chain of the neighbor-coupled harmonic oscillators
and others~\cite{Lee,Lee1} as well as in collective particle dynamics in simple liquids~\cite{Mokshin_TMP_2015}, where the TCF of the local density fluctuations is considered.

The three considered cases allow one to find solution of Eq.
(\ref{first}) at the three different values of the memory parameter:
$\delta=1$ with solution (\ref{bessel}); the memory-free case with solution
(\ref{expon}) corresponds to $\delta \to \infty$, whereas
solution (\ref{cos}) was obtained for the system with ideal memory
at $\delta \to 0$. So, if we shall generalize Eqs.~(\ref{expon}) and
(\ref{cos}) in a unified functional dependence, then we can obtain the
following solution of Eq.~(\ref{first}) in terms of the
Mittag-Leffler function~\cite{tranc}:
\begin{equation}\label{polin1}
  M_{0}(t)=\sum_{k=0}^{\infty}\frac{
(-\Omega_{1}^{2}\tau_{1}^{1-\nu}t^{\nu+1})^{k}}{\Gamma(\nu
k+k+1)},
\end{equation}
where $\Gamma(\ldots)$ is the Gamma function, $\tau_1$ is the time scale of the memory function $M_1(t)$, the frequency parameter is determined by (\ref{eq: freq_par}), and
the  dimensionless parameter $\nu$ is the redefined memory measure:
\begin{equation}\label{eq: delta}
  \nu=(2/\pi)\arctan(1/\delta),
\end{equation}
and $\nu\in[0;1]$.
For an strong memory case, i.e. $\delta \to 0$, Eq.~(\ref{polin1})
gives expansion in series of Eq.~(\ref{cos}), whereas for a memory-free limit
with $\delta \to \infty$ we obtain expansion of Eq.~(\ref{expon}).
Relation~(\ref{polin1}) has a stretched
exponential behavior at short times and demonstrate an inverse power-law
relaxation at long times.
It should be noted that the similar solution of
integro-differential equations was proposed earlier by Stanislavskii in Ref. \cite{Stanislavskii} on the basis
of applying the fractional calculus technique.

Moreover, one can obtain the general solution of Eq.~(\ref{first}) by interpolation of all three solutions (\ref{expon}), (\ref{cos}) and
(\ref{bessel}). Assuming the smooth parabolic crossover from the strong memory and memory-free limits to a case of the moderate memory ($\delta =1$) we
obtain
\begin{eqnarray}\label{polin2}
  M_{0}(t)&=&4\left(\nu-\frac{1}{2}\right)^{2}\sum_{k=0}^{\infty}
\frac{(-\Omega_{1}^{2}\tau_{1}^{1-\nu}t^{\nu+1})^{k}}{\Gamma(\nu
k+k+1)}\nonumber \\
&+&\left [1-4\left(\nu-\frac{1}{2}\right)^{2}\right
]\sum_{k=0}^{\infty} \frac{(-2 \Omega_{1}^{2}t^{2})^{k}}{k!(k+1)!},
\end{eqnarray}
where the parameter $\nu$ is defined by (\ref{eq: delta}). The
significance of the first contribution increases at approaching
$\delta$ to the zeroth value or at $\delta \to \infty$. Thus, for
example, Eq.~(\ref{polin2}) gives a standard exponential relaxation
in the memory-free limit with $\delta \rightarrow \infty$. The
second contribution in (\ref{polin2}) provides the Gaussian behavior
for the short-time range $t<\Omega_{1}^{-1}$. Further, the numerical
coefficient before sum in the second contribution of
Eq.~(\ref{polin2}) dominates in the intermediate region, where the
time scales of memory function and of the VACF are comparable. This
contribution becomes maximal at $\delta=1$, whereas the first item
turns into zero~\footnote{Notice that the second contribution in
expression (\ref{polin2}) can be considered as a particular case of
the Mainardi function \cite{Mainardi} (or the Wright function
\cite{tranc}), which includes such functions as the Gaussian
function, the Dirac delta-function and others.}.

\begin{figure}[h!]
\leavevmode
\centering
\includegraphics[width=3.4in, angle=0]{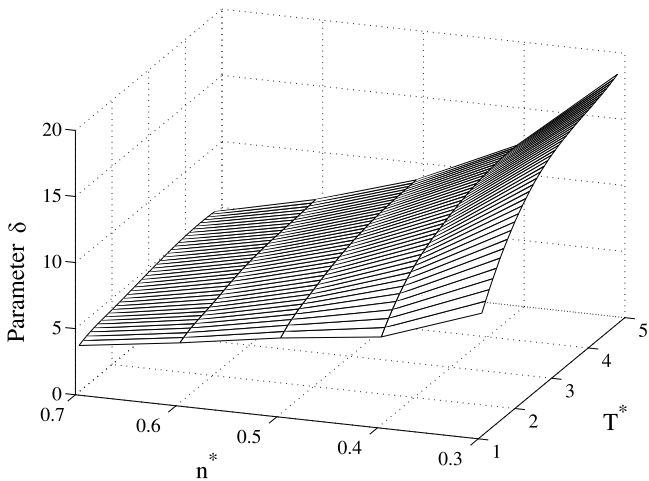}
\caption{Density- and temperature-dependence of the
memory parameter $\delta$ computed on the basis of molecular dynamics simulation results for the VACF of the Lennard-Jones
liquid for the following region of the ($n^*,T^*$)-phase diagram: $0.3\leq n^*\leq
0.7$ and $1\leq T^*\leq 4.7$, $n^*=n\sigma^3$, $T^*=k_B
T/\epsilon$, where $\sigma$ and $\epsilon$ are the parameters of the Lennard-Jones
potential \cite{my}. \label{fig: memory}}
\end{figure}

Both the memory-free situation with absolutely uncorrelated particle motions and the strong-memory case with the pronounced correlations in the particle velocities related with the vibrational particle dynamics are only limit ones for a real liquids. A real liquid (a fluid) tends to the first one at high temperatures and
low density, whereas it approaches to another limit at low temperatures with large values of the density. Moreover, it is realized a regime of dense fluids, where the surrounding medium with neighbor
particles has an appreciable impact on a forward moving particle
($\alpha$), causing the so-called vortex diffusion
\cite{Cohen} and existence of the power law decay of $M_{0}(t)$
with time. This indicates on the memory effects in single-particle dynamics, albeit the memory is far to be strong.
Our numerical estimations
of the memory effects for self-diffusion processes in the Lennard-Jones fluids
\cite{my} has found that the parameter $\delta$
at the reduced temperature $T^*\approx 1$ and the reduced density
$\ n^*=0.5$ has a value $\approx 5.9$, and then it increases with
the growth of temperature and the decreasing of density. The parameter $\delta$
achieves value $\approx 8.7$ at the temperature $T^*\approx 4.8$
and the density $n^*=0.5$, and demonstrates non-linear smooth
Markovization. For more visibility of aforesaid we present in
Fig.~(\ref{fig: memory}) the density- and  the temperature-dependence of the
memory parameter $\delta$ calculated for the VACF of
Lennard-Jones fluids \cite{my}.

Moreover, as known, the Bessel function of Eq.~(\ref{bessel}) has the
following asymptotic behavior \cite{Abram}:
\[
J_{1}(z)=\sqrt{2/\pi
z}\left \{ \cos \left ( z-(3\pi/4) \right )+
\mathcal{O}(|z|^{-1})\right \}.
\]
Then, returning to Eq.~(\ref{polin2}), it is easy to make sure that in a case of
moderate memory this equation yields the following long-time tail:
\begin{equation}\label{long}
  M_{0}(t) \propto t^{-3/2},
\end{equation}
which is a well-known feature of the VACF's  of liquids~\cite{Alder,Levesque}.
The long-time tails of the VACF of a simple liquid can be reproduced within the microscopic \textit{mode coupling theories}
(see, for example, \cite{Hansen}), according to which where it is related to a viscous  mode.
The approach presented in this study  is consistent with the mode-coupling theories and provides a theoretical description, in which information about complex correlated vibrational-diffusive motions of the particles is included into a single parameter $\delta$.
\begin{figure}[h!]
\leavevmode
\centering
\includegraphics[width=3.4in, angle=0]{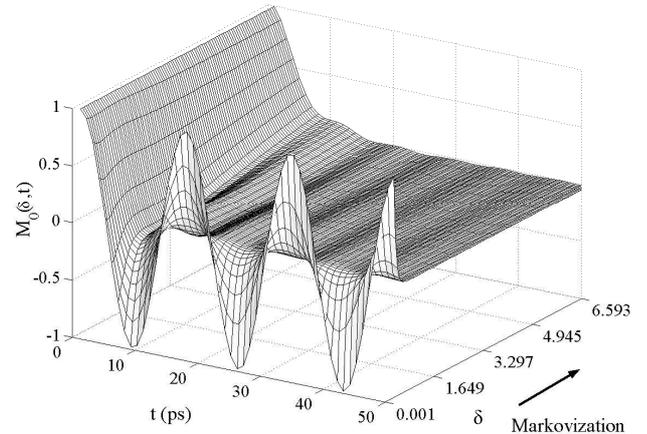}
\caption{Velocity autocorrelation function $M_{0}(\delta,t)$ as a
function of the time $t$ and of the memory parameter $\delta$
computed from Eq.~(\ref{polin2}) for model case:  $\delta \in
[0.001;\;6.593]$, where $\Omega_{1}^{2}=5\,$ps$^{-2}$ and the time
scale $\tau_{1}=1/\sqrt{2\Omega_{1}^{2}\delta}$. The arrow points to
the direction of Markovization.}
\end{figure}
Furthermore, it is seen that Eqs.~(\ref{polin1}) and (\ref{polin2}) contain
such the characteristics of a many-particle system as the characteristic time scale of the memory $\tau_1$ and the averaged frequency $\Omega_{1}^{2}$, which is defined through the radial distribution function $g(r)$ and the particle's interaction potential $U(r)$. These
quantities can be calculated from their definitions for concrete
systems, or may be taken from molecular dynamics simulations (see,
for example, \cite{Tuckerman}). On the other hand, while analysing
experimental data, the term $\delta$ may be used as a
fitting parameter to do the quantitative estimation of the memory effects in the considered system.

As an example, we demonstrate in Fig.~2 results of Eq.~(\ref{polin2}) for a model system with the frequency parameter $\Omega_{1}^{2}=5$~ps$^{-2}$. The memory parameter $\delta$ varies
within the interval from $0.001$ to $6.593$, whereas the memory time scale $\tau_1$  has been defined here as
$\tau_{1}=1/\sqrt{2 \Omega_{1}^{2}\delta}$. Thus, the presented
results include situations of a system with a strong memory when the
ratio between the time scales of the VACF and of its memory function
archive the value $0.001$, and situations when these time scales are
comparable. One can see from this figure that the
oscillations in the VACF disappear with attenuation of the memory effects (at
Markovization)~\cite{my}. These oscillations will disappear completely at
$\delta \rightarrow \infty$.  The stronger
memory effects in the single-particle dynamics of a system, the more considerable the amplitude of
fluctuations and their decay.

\begin{figure}[h!]
\leavevmode
\centering
\includegraphics[width=3.0in, angle=0]{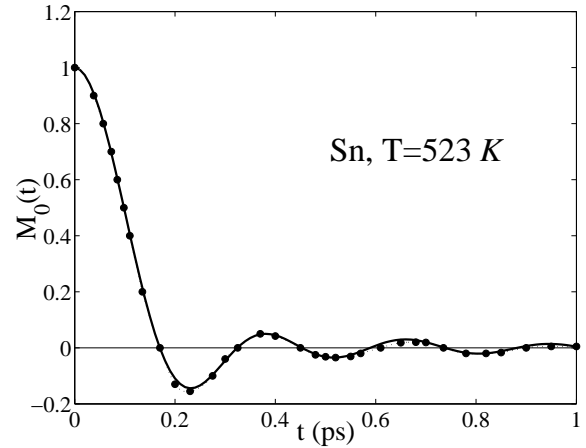}
\caption{VACF of liquid tin at the temperature $T=523$~K (the melting temperature is $T_m=505.08$~K) calculated from Eq.
(\ref{polin2}) with parameters $\Omega_{1}^{2} \simeq 130$ ps$^{-2}$,
$\delta \simeq 0.75$ and
$\tau_{1}=1/\sqrt{2 \Omega_{1}^{2}\delta}$ (solid curve) and
obtained from the classical molecular dynamics simulations~\cite{jap1} ($\bullet
\bullet \bullet)$. The value of the memory parameter confirms
the presence of strong memory effects in the single-particle dynamics, which appear due to the specific ion-ion interaction and the
correlated disorder in the structure of the melting metal. }
\end{figure}
The dependence presented by Eq.~(\ref{polin2}) is supported by
experimental results. Experimental and molecular dynamic studies of
simple liquids such as liquid tin \cite{jap1}, liquid germanium and
lithium \cite{jap2}, liquid selenium \cite{jap3}, liquid sodium
\cite{jap4}, Lennard-Jones fluids \cite{jap5} and other systems
\cite{Tuckerman,Morgado} have allowed one to discover the relaxation
of the VACF with the signatures of the pronounced memory effects,
which is manifested, in particular, in algebraic decay of the VACF.
To test Eq.~(\ref{polin2}) for the liquids, we perform  the
compuations for the cases liquid tin and liquid lithium, for which
the VACF's were found before from molecular dynamics
simulations~\cite{jap1,jap2}. In Figs.~3 and 4, numerical solutions
of Eq.~(\ref{polin2}) are compared with the results of molecular
dynamics simulations~ \cite{jap1,jap2}. As for Fig. 3, the full
circles represent the VACF of liquid tin at $T=523$~K (the melting
temperature $T_{m}=505.08$~K) calculated by the classical molecular
dynamics \cite{jap1} and the solid curve shows  solution of the GLE
(\ref{polin2}) with $\Omega_{1}^{2}\simeq 130$~ps$^{-2}$, $\delta
\simeq 0.75$ and $\tau_{1}=1/\sqrt{2 \Omega_{1}^{2}\delta}$.
\begin{figure}[h!]
\leavevmode
\centering
\includegraphics[width=3.0in, angle=0]{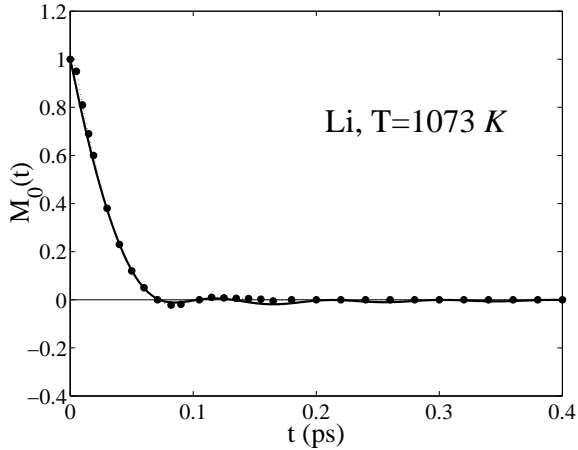}
\caption{VACF of liquid lithium at the temperature $T=1073$~K (the
melting temperature is $T_m =453.65$~K) found from
Eq.~(\ref{polin2}) with the parameters $\Omega_{1}^{2}\simeq
120\,$ps$^{-2}$, $\delta \simeq 10$ and
$\tau_{1}=1/\sqrt{2\Omega_{1}^{2}\delta}$ (solid curve) and from
molecular dynamics simulations~\cite{jap2} ($\bullet \bullet
\bullet)$. The value of the memory parameter $\delta$ reveals a weak
memory in single-particle dynamics in liquid lithium for this
thermodynamic state, that can be related with a weak influence of
the ``cage effects''~\cite{Hansen} as well as with a low ordering in
the system. As a result, a tagged diffusing atom can propagate over
larger distances without collisions with neighboring atoms.}
\end{figure}
It is seen that Eq.~(\ref{polin2}) well agrees with the molecular
dynamics simulations over the whole time interval. The value of the
memory parameter reveals the pronounced memory effects in
self-diffusion phenomena in liquid tin near its melting point.
Strong memory effects, which take place in single-particle dynamics
of liquid near its melting point, can be related with the structural
transformations of the system~\cite{Khusnutdinoff/Mokshin_2010}.
Fig. 4 shows the VACF of liquid lithium at $T=1073$~K (the melting
temperature $T_{m}=453.65$~K) determined from the molecular dynamics
simulations~\cite{jap2} as full circles, whereas the solid line
corresponds to the solution of Eq.~(\ref{polin2}) with the frequency
parameter $\Omega_{1}^{2}\simeq 120\,$ps$^{-1}$, the memory
parameter $\delta \simeq 10$ and the time scale of the memory
function $\tau_{1}=1/\sqrt{2 \Omega_{1}^{2}\delta}$. As may be seen
from Fig. 4, the theoretical results and the data of the molecular
dynamics simulations~\cite{jap2} are in good agreement. The VACF of
liquid lithium at this temperature does not practically oscillate.
This is direct indications of weak memory effects in the system,
that can be caused by the absence of structural order in the the
system and by the diffusive character of the single-particle
dynamics~\cite{Mokshin/Galimzyanov_JPCB_2013}.

Finally, the presented approach may also be used to investigate
microscopical dynamics in more complex liquids, whose interatomic
potentials include angular-dependent contributions. The memory
function $M_{1}(t)$ of these systems, which represents the TCF of
the stochastic force $f(t)$, will represent a combination of certain
relaxation coupling modes. As a result, the VACF will have a more
complex time behavior.

\vskip 1cm

We thank M. Howard Lee for useful discussions. This work was partially supported by the subsidy allocated to Kazan Federal University for the state assignment in the sphere of scientific activities.

\end{document}